\documentclass[12pt]{article}
\usepackage{amsthm,amssymb,amsmath}
\usepackage[american,british]{babel}

\newtheorem*{pro}{Proposition}

\newtheorem{lem}{Lemma}

\newcommand{\dif}{\operatorname{d}}
\renewcommand{\t}{{\operatorname{t}}}

\renewcommand{\L}{\operatorname{L}}
\newcommand{\diag}{\operatorname{diag}}
\newcommand{\bX}{{\boldsymbol X}}
\newcommand{\ba}{{\boldsymbol a}}

\newcommand{\bx}{{\boldsymbol x}}
\newcommand{\be}{{\boldsymbol e}}
\newcommand{\bc}{{\boldsymbol c}}
\newcommand{\bu}{{\boldsymbol u}}

\newcommand{\bxi}{{\boldsymbol \xi}}
\newcommand{\bzeta}{{\boldsymbol \zeta}}

\newcommand{\R}{{\mathbb R}}

\newcommand{\D}{{\partial}}

\begin{document}

\title{Reduced Vectorial Ribaucour Transformation\\
for the  Darboux-Egoroff Equations}

\author{Q. P. Liu$^1$\thanks{On leave of absence from
Beijing Graduate School, CUMT, Beijing 100083, China}
\thanks{Supported by {\em Beca para estancias temporales
de doctores y tecn\'ologos extranjeros en
Espa\~na: SB95-A01722297}}
   $\,$ and Manuel Ma\~nas$^{2,1}$\thanks{Partially supported by CICYT:
 proyecto PB95--0401}\\
$^1$Departamento de F\'\i sica Te\'orica II,\\ Universidad
Complutense,\\ E28040-Madrid, Spain.\\ $^2$Departamento de
Matem\'atica Aplicada y Estad\'\i stica,\\ Escuela Universitaria de
Ingenier\'\i{}a T\'ecnica Areona\'utica,\\ Universidad
Polit\'ecnica de Madrid,\\ E28040-Madrid, Spain.}
\date{}

\maketitle

\begin{abstract}
The vectorial fundamental transformation for the Darboux equations
is reduced to the symmetric case. This is combined with the
orthogonal reduction of Lam\'{e} type  to obtain  those  vectorial
Ribaucour transformations  which preserve the Egoroff reduction. We
also show that a permutability property holds for all these
transformations. Finally, as an example, we apply these
transformations to the Cartesian background.
\end{abstract}
\newpage

\section{Introduction}

At the turn of this century a number of results on differential
geometry were already well established
\cite{Darboux1,Darboux2,Eisenhart1,bianchi}. We are thinking of
conjugate nets described by the Darboux equations
\cite{Darboux1,Eisenhart1} and related transformations
\cite{Eisenhart2} of Laplace, L\'{e}vy  \cite{levy} and fundamental
\cite{jonas,Eisenhart3} type; and  the orthogonal nets, described
by the Lam\'{e} equations \cite{lame,Darboux2},   and their Ribaucour
transformations \cite{Ribaucour,Darboux2}. The Lam\'{e} equations
describe flat diagonal metrics, among which we find a distinguished
class: those of Egoroff type. These particular classes of flat
diagonal metrics are described by the Darboux-Egoroff equations
\cite{Darboux2}, that were first proposed by Darboux
\cite{Darboux3} and studied further in
\cite{Ribaucour2,petot,Fouche} and finally, as it was recognized by
Darboux  \cite{Darboux2}, Egoroff gave an almost definitive
treatment in \cite{Egorov}.

The mentioned results are deeply connected with the modern theory
of integrable systems. It is well known that integrable equations
like Liouville or sine-Gordon were first considered in the context
of differential geometry: minimal and pseudo-spherical surfaces. It
has been recently discovered that  the $N$-component
Kadomtsev-Petviashvili (KP) hierarchy \cite{kp} describes the
iso-conjugate transformations of the Darboux equations, and its
vertex operators correspond precisely to Laplace, L\'{e}vy and
fundamental transformations \cite{dmmms}. Moreover, the
$N$-component BKP hierarchy \cite{bkp} models the iso-orthogonal
deformations of the Lam\'{e} equations, being its vertex operator the
Ribaucour transformation \cite{mm}. Hence, one easily concludes the
strong relation between Soliton Theory and Classical Differential
Geometry. Let us mention that discrete integrable systems have the
same connection with  modern discrete geometry \cite{dsm}.

Recently in \cite{Dubrovin} it was shown that the Egoroff metrics
play a fundamental role in the classification of two dimensional
topological conformal field theories. The partition function of the
deformed semisimple theories satisfies certain associativity or
Witten-Dijkgraaf-Verlinde-Verlinde equations, which are connected
with an Egoroff metric.

The integration of the Darboux equations was performed in \cite{zm}
through the $\bar\partial$-method, while very recently the Lam\'{e}
equations have been integrated within the inverse scattering
technique in \cite{z} and by the algebro-geometrical approach in
\cite{Krichever}.

In previous papers we have considered the iteration of the
mentioned transformations within the modern soliton theory: in
\cite{lm1} we studied the L\'{e}vy transformation and in \cite{lm} the
Ribaucour transformation, using in this latter a vectorial
approach. We want to reduce the vectorial fundamental
transformation \cite{mds,dsm} to the symmetric case, and further
the vectorial Ribaucour transformation to the Darboux-Egoroff
equations.

The layout out of this paper is as follows: in \S 2 we consider the
symmetric reduction of the Darboux equations obtaining the
vectorial symmetric fundamental transformations for all the
geometrical data; next, in \S 3 we combine these results with the
ones of \cite{lm} to get the reduction of the vectorial Ribaucour
to the Darboux-Egoroff equations, giving the expressions of all the
transformed relevant geometrical data. In both sections we consider
the dressing of the Cartesian background and prove that the
permutability property is preserved under all the reductions
considered. We must recall the reader that the permutability
property of these transformations is an important issue that for
the fundamental transformation was considered in
\cite{demoulin,dsm} and for the Ribaucour transformation in
\cite{bianchi1,bianchi,lm}.

\section{The vectorial fundamental transformation for the
symmetric Darboux equations}
 The Darboux equations
\begin{equation} \label{dar}
\frac{\partial\beta_{ij}}{\D u_k}-\beta_{ik}\beta_{kj}=0,
\;\;
i,j,k=1,\dotsc, N,\;
\text{with $i,j,k$ different},\end{equation}
for the $N(N-1)$ functions
$\{\beta_{ij}\}_{\substack{i,j=1,\dotsc,N\\i\neq j}}$ of
$\bu:=(u_1,\dotsc,u_N)$, characterize $N$-dimen\-sional
submanifolds of $\R^D$, $N\leq D$, parametrized by conjugate
coordinate systems \cite{Darboux1,Eisenhart1}, and are the
compatibility conditions of the following linear system
\begin{equation} \label{X}
\frac{\partial \bX_j}{\D u_i} = \beta_{ji} \bX_i, \quad i,j=1,\dotsc,N,\quad
i\ne j,
\end{equation}
involving suitable $D$-dimensional vectors $\bX_i$, tangent to the
coordinate lines.

 The so called Lam\'{e} coefficients satisfy
\begin{equation} \label{H}
\frac{\partial H_j}{\D u_i} = \beta_{ij} H_i, \quad i,j=1,\dotsc,N,\quad
i\ne j,
\end{equation}
and the points of the surface $\bx=(x_1,\dots,x_N)$ can be found by
means of
\begin{equation}\label{points}
\frac{\D \bx}{\D u_i}= \bX_i H_i,\quad i=1,\dotsc, N,
\end{equation}
which is equivalent to the  Laplace equation
\[
\frac{\D^2\bx}{\D u_i\D u_j}=\frac{\D \ln H_i}{\D u_j}\frac{\D \bx}{\D u_i}
+\frac{\D \ln H_j}{\D u_i}\frac{\D \bx}{\D u_j},
\quad i,j=1,\dotsc,N,\;\; i\neq j.
\]

The fundamental transformation for the Darboux system was
introduced in \cite{jonas,Eisenhart3},
and its vectorial extension was given
in \cite{mds,dsm}. It requires the introduction of a potential in
the following manner: given vector solutions $\bxi_i\in V$ and
$\bzeta_i^*\in W^*$ of \eqref{X} and \eqref{H}, $i=1,\dotsc,N$,
 respectively, where
$V,W$ are linear spaces and $W^*$ is the dual space of $W$,
one can define a potential matrix
$\Omega(\bxi,\bzeta^*):W\to V$ through the equations
\begin{equation}\label{potential}
\frac{\D   \Omega(\bxi,\bzeta^*)}{\D u_i}=\bxi_i\otimes\bzeta^*_i.
\end{equation}
\newtheorem*{VF}{\textit{Vectorial Fundamental Transformation}}

\begin{VF}
Given solutions $\bxi_i\in V$ and $\bxi_i^*\in V^*$ of
\eqref{X} and \eqref{H}, $i=1,\dotsc,N$,
 respectively,
new rotation coefficients $\hat\beta_{ij}$, tangent vectors
$\hat{\bX}_i$, Lam\'e coefficients $\hat H_i$ and points of the
surface $\hat{\bx}$ are given by
\begin{equation}\label{vecfun}
\begin{aligned}
\hat\beta_{ij}&=\beta_{ij}-
\langle\bxi^*_j, \Omega(\bxi,\bxi^*)^{-1}\bxi_i\rangle,\\
\hat{\bX}_i&=\bX_i-\Omega(\bX,\bxi^*)\Omega(\bxi,\bxi^*)^{-1}\bxi_i,\\
\hat H_i&=H_i-\bxi_i^*\Omega(\bxi,\bxi^*)^{-1}\Omega(\bxi,H),\\
\hat{\bx}&=\bx-\Omega(\bX,\bxi^*)\Omega(\bxi,\bxi^*)^{-1}\Omega(\bxi,H).
\end{aligned}
\end{equation}
\end{VF}
\noindent
Here we are assuming that $\Omega(\bxi,\bxi^*)$ is invertible. We
shall refer to this transformation as vectorial fundamental
transformation with transformation data $(V,\bxi_i,\bxi_i^*)$.

The symmetric reduction requests the  rotation coefficients to be
symmetric; i. e.,
\begin{equation}
\label{symmetric}
\begin{gathered}
\frac{\partial\beta_{ij}}{\D u_k}-\beta_{ik}\beta_{kj}=0,
\;\;
i,j,k=1,\dotsc, N,\;
\text{with $i,j,k$ different},\\
\beta_{ij}-\beta_{ji}=0,\quad i,j=1,\dots,N,\;\; i\neq j.
\end{gathered}
\end{equation}
This implies  local existence of  potentials $V$ and $\Phi$ such
that $|\bX_i|^2=\dfrac{\partial V}{\D u_i}$ and
 \begin{equation}\label{sh}
 H_i^2=\dfrac{\partial\Phi}{\D u_i},\quad i=1,\dots,N.
\end{equation}

We now consider which transformation data $(V,\bxi_i,\bxi_i^*)$
gives a vectorial fundamental transformation that preserves the
symmetric Darboux equations  \eqref{symmetric}.

We have the following observations:
\begin{lem}
\begin{enumerate}
\item
Given a solution $\bxi_i\in V$ of \eqref{X} then
\begin{equation}\label{*}
\bxi^*_i:=\bxi_i^{\t}L
\end{equation}
where $^\t$ means transpose and $L\in\L(V)$ is a linear operator on
$V$, is a $V^*$-valued solution of
\eqref{H} if and only if \eqref{symmetric} holds.
We shall say that $L$ is the associated linear operator.
 \item
 Given symmetric $\beta$'s, $\bxi_i\in V$ and $\bzeta_i\in W$ solutions of
\eqref{X} and $\bxi^*_i$ and $\bzeta^*_i$ as prescribed in \eqref{*},
$i=1,\dotsc,N$, with associated linear operators $L$ and $M$,
respectively; then:
\[
\frac{\D}{\D u_i}(L^\t\Omega(\bxi,\bzeta^*)-\Omega(\bzeta,\bxi^*)^{\t}M)=0,\quad
i=1,\dotsc,N.
\]
\item Suppose given a solution $\beta_{ij}$ of the symmetric Darboux
equations
\eqref{symmetric},  $\bxi_i\in V$ and $\bzeta_i\in W$ solving
\eqref{X} and  $\bxi_i^*$ and $\bzeta_i^*$ as prescribed in
\eqref{*}, with associated linear operators $L$ and $M$, respectively. Then,
if
\begin{equation}\label{constraints}
\begin{aligned}
L^\t\Omega(\bxi,\bzeta^*)-\Omega(\bzeta,\bxi^*)^\t M&=0,\\
L^\t\Omega(\bxi,\bxi^*)-\Omega(\bxi,\bxi^*)^\t L&=0,
\end{aligned}
\end{equation}
 the vectorial fundamental transformation \eqref{vecfun}:
\begin{align*}
\hat\beta_{ij}&=\beta_{ij}-
\langle\bxi^*_j, \Omega(\bxi,\bxi^*)^{-1}\bxi_i\rangle,\\
\hat{\bzeta}_i&=\bzeta_i-\Omega(\bzeta,\bxi^*)\Omega(\bxi,\bxi^*)^{-1}\bxi_i,\\
\hat\bzeta_i^*&=\bzeta_i^*-\bxi_i^*\Omega(\bxi,\bxi^*)^{-1}\Omega(\bxi,\bzeta^*),
\end{align*}
is such that
\[
\hat\bzeta^*_i:=\hat\bzeta_i^\t M.
\]
\end{enumerate}
\end{lem}
\begin{proof}
Point 1 is trivial to check; for 2, using \eqref{potential} and the
definition
\eqref{*} we have
\[
\frac{\D}{\D u_i}(L^\t\Omega(\bxi,\bzeta^*)-\Omega(\bzeta,\bxi^*)^\t M)
=L^\t\bxi_i\otimes\bzeta_i^\t M-L^\t\bxi_i\otimes\bzeta_i^\t M=0.
\]
For 3, using \eqref{potential}, \eqref{vecfun} and \eqref{*} we
find that
\[
\hat\bzeta_i^*=\bzeta_i^\t M-\bxi_i^\t L\Omega(\bxi,\bxi^*)^{-1}
\Omega(\bxi,\bzeta^*),
\]
and the constraints \eqref{constraints}  gives
\[
\hat\bzeta_i^*=\bzeta_i^\t M-\bxi_i^\t\big(\Omega(\bxi,\bxi^*)^\t\big)^{-1}
L^\t\Omega(\bxi,\bzeta^*)=\Big[
\bzeta_i-\Omega(\bxi,\bzeta^*)\Omega(\bxi,\bxi^*)^{-1}
\bxi_i
\Big]^\t M,
\]
that recalling \eqref{vecfun} implies the statement.
\end{proof}

Therefore, as $\Omega(\bxi,\bzeta^*)$ and $\Omega(\bzeta,\bxi^*)$
are defined by \eqref{potential} up to additive constant matrices,
 we may take them such that
\begin{equation*}
L^\t\Omega(\bxi,\bzeta^*)-\Omega(\bzeta,\bxi^*)^\t M=0.
\end{equation*}

These  lemmas imply:
\newtheorem*{VSF}{\textit{\textbf{Vectorial Symmetric Fundamental Transformation}}}

\begin{VSF}
The vectorial fundamental transformation \eqref{vecfun} preserves
the symmetric Darboux equations  \eqref{symmetric} whenever the
transformation data ($V$, $\bxi_i$, $\bxi_i^*$) satisfies
\begin{align*}
&\bxi_i^*=\bxi_i^\t L,\\
 &L^\t\Omega(\bxi,\bxi^*)-\Omega(\bxi,\bxi^*)^\t L=0.
\end{align*}
\end{VSF}

We say that $(V,\bxi_i,L)$ is the transformation data for this
particular vectorial fundamental transformation that we shall call
vectorial symmetric fundamental transformation.

\subsection{Permutability of vectorial symmetric fundamental transformations}
The vectorial fundamental transformations permute among them
\cite{dsm}:
\newtheorem*{Per}{\textit{Permutability of Vectorial Fundamental
Transformations}}
\begin{Per}
The vectorial fundamental transformation with transformation data
\[
\bigg(V_1\oplus V_2,
\begin{pmatrix} \bxi_{i,(1)}\\ \bxi_{i,(2)}\end{pmatrix},
(\bxi^*_{i,(1)},\bxi^*_{i,(2)})\bigg)
\]
coincides with the following
composition of vectorial fundamental transformations:
\begin{enumerate}
\item First transform with data
\[
(V_2,\bxi_{i,(2)},\bxi_{i,(2)}),
\]
 and denote the transformation by $^\prime$.
\item On the result of this transformation apply a second one with data
 \[
(V_1, \bxi_{i,(1)}^\prime, \bxi_{i,(1)}^\prime).
\]
\end{enumerate}
\end{Per}

Therefore, the composition of two vectorial fundamental
 transformations yields a
new vectorial fundamental transformation. When these two
transformations are done in different order  the resulting composed
vectorial fundamental transformation is equivalent, through
conjugation by a permutation matrix, to the first composed
vectorial fundamental transformation, so that all the geometrical
data are identical for both composed transformations; hence, the
permutability character of these transformations. Moreover,  it
also follows that the vectorial fundamental transformation is just
a superposition of a number of fundamental transformations.

One can easily conclude that this result can be extended to the
vectorial symmetric fundamental transformation:

\begin{pro}
The vectorial symmetric fundamental transformation with
transformation data
\[
\bigg(V_1\oplus V_2,
\begin{pmatrix} \bxi_{i,(1)}\\ \bxi_{i,(2)}\end{pmatrix},
\begin{pmatrix} L_{(1)}&0\\0& L_{(2)}\end{pmatrix}
\bigg),
\]
 coincides with the following
composition of vectorial symmetric fundamental transformations:
\begin{enumerate}
\item First transform with data
\[
(V_2,\bxi_{i,(2)},L_{(2)}),
\]
 and denote the transformation by $^\prime$.
\item On the result of this transformation apply a second one with data
   \[
(V_1, \bxi_{i,(1)}^\prime,L_{(2)}).
\]
\end{enumerate}
\end{pro}

\begin{proof}
Because the transformation data follows the prescription of our Theorem
they must satisfy
\begin{align*}
&(\bxi_{i,(s)}^*)^\t=L_{(s)}\bxi_{i,(s)},\quad s=1,2\\
&L_{(s)}^\t\Omega(\bxi_{(s)},\bxi_{(s)}^*)-\Omega(\bxi_{(s)},\bxi_{(s)}^*)^\t
L_{(s)}=0,\quad s=1,2,\\
&L_{(1)}^\t\Omega(\bxi_{(1)},\bxi_{(2)}^*)-
\Omega(\bxi_{(2)},\bxi_{(1)}^*)^\t L_{(2)}=0.
\end{align*}
The first vectorial fundamental transformation is a vectorial
symmetric one with data $(V,\bxi_{i,(2)},L_{(2)})$. Lemma 3 implies
that the vectorial fundamental transformation of point 2 is also a
vectorial symmetric fundamental transformation.
\end{proof}

Form these results we conclude that the composition of scalar
symmetric fundamental transformations results in a vectorial
symmetric fundamental transformation with associated matrix of
diagonal type. In fact, when the associated matrix $L$ is not
diagonal the corresponding transformation can not be obtained by
means of composition only, we need also a suitable coalescence of
eigen-values.
\subsection{Dressing the Cartesian background}

The Cartesian net  has $\bX_i=\be_i$, being
$\{\be_i\}_{i=1,\dotsc,N}$ a linear independent set of vectors  of
$\R^D$, $H_i=1$, the coordinates are $\bx(\bu)=\bu$ and vanishing
rotation coefficients $\beta_{ij}=0$. Hence,
\[
\bxi_i=\bxi_i(u_i)\in\R^M,
\]
and
\[
\Omega(\bxi,\bxi^*)(\bu)=\sum_{k=1,\dotsc, N}\Omega_i(u_i)
\]
 with
\begin{align*}
&
\Omega_i(u_i)=\int_{u_{i,0}}^{u_i}\dif u_i\,\bxi_i\otimes
\bxi_i^\t L+\Omega_{i,0} ,\\
&L^\t\Omega_{i,0}-\Omega_{i,0}^\t L=0.
\end{align*}
Thus
\begin{align*}
\Omega(\bX,\bxi^*)(\bu)&=A+
\sum_{i=1,\dotsc, N} \be_i\otimes\int_{u_{i,0}}^{u_i}
\dif u_i\, \bxi_i^\t(u_i)L,\\
\Omega(\bxi,H)(\bu)&=\bc+\sum_{i=1,\dotsc, N} \int_{u_{i,0}}^{u_i}
\dif u_i\, \bxi_i(u_i),
\end{align*}
where $A$ is a constant $D\times M$ matrix and
 $\bc\in\R^M$ is a constant vector, and the
symmetric conjugate net is given by
\begin{multline*}
\bx(\bu)=\bu-
\bigg[A+\sum_{i=1,\dotsc, N} \be_i\otimes
\int_{u_{i,0}}^{u_i}\dif u_i\,\bxi_i^\t(u_i)\bigg]
\bigg[\sum_{i=1,\dotsc, N}\Omega_i(u_i)\bigg]^{-1} \\
\times\bigg[\bc+
\sum_{i=1,\dotsc, N} \int_{u_{i,0}}^{u_i}
\dif u_i \bxi_i(u_i)\bigg].
\end{multline*}

\section{The vectorial fundamental transformation for the
Egoroff metrics}

The Lam\'{e} equations describe $N$-dimensional conjugate orthogonal
systems of coordinates \cite{lame,Darboux2,tsarev}:
\begin{align}
&\frac{\partial\beta_{ij}}{\D u_k}-\beta_{ik}\beta_{kj}=0,\;\;
i,j,k=1,\dotsc, N,\;
\text{with $i,j,k$ different},\label{lame1}\\
&\frac{\partial\beta_{ij}}{\D
u_i}+
\frac{\partial\beta_{ji}}{\D u_j}+
\sum_{\substack{k=1,\dotsc,N\\ k\neq i,j}}
\beta_{ki}\beta_{kj}=0,\quad i,j=1,\dotsc,N,\;
i\neq j.\label{lame2}
\end{align}
Now we have orthogonal tangent directions,
$\bX_i\cdot\bX_j=\delta_{ij}$. In fact, if $N=D$, the Lam\'{e}
coefficients $H_i$ allows us to construct a flat diagonal metric:
\begin{equation}\label{metric}
\dif s^2=\sum_{i=1}^N H_i(\bu)^2 \dif u_i\otimes\dif u_i,
\end{equation}
for which $x_1,\dots, x_N$ are flat coordinates.
 The reduction of the vectorial fundamental transformation to the
 orthogonal case; i. e., the vectorial Ribaucour transformation,
 was studied by us in \cite{lm}. The main results
 of it are
\begin{lem}
\begin{enumerate}
\item Given a solution $\bxi_i\in V$ of \eqref{X} then
\begin{equation}\label{*l}
\bxi^*_i:=\Big(\frac{\D\bxi_i}{\D u_i}+
\sum_{\substack{k=1,\dotsc,N\\ k\neq i}}\bxi_k \beta_{ki}\Big)^{\t},
\end{equation}
 is a $V^*$-valued solution of \eqref{H} if and only if \eqref{lame2} holds.
\item
 Given $\beta$'s solving the Lam\'e
equations \eqref{lame1} and \eqref{lame2}, $\bxi_i\in V$ and
$\bzeta_i\in W$ solutions of
\eqref{X} and $\bxi^*_i$ and $\bzeta^*_i$ as prescribed in \eqref{*l},
$i=1,\dotsc,N$, then:
\[
\frac{\D}{\D u_i}\Big(\Omega(\bxi,\bzeta^*)+\Omega(\bzeta,\bxi^*)^\t-
\sum_{k=1,\dotsc,N}\bxi_k\otimes\bzeta_k^\t\Big)=0,\quad i=1,\dotsc,N,
\]
\item Suppose given a solution $\beta_{ij}$ of the Lam\'e equations
\eqref{lame1} and \eqref{lame2},  $\bxi_i\in V$ and $\bzeta_i\in W$
solving \eqref{X} and  $\bxi_i^*$ and $\bzeta_i^*$ as prescribed in
\eqref{*l}. Then,
if
\begin{equation}\label{constraintsl}
\begin{aligned}
\Omega(\bxi,\bzeta^*)+\Omega(\bzeta,\bxi^*)^\t&=
\sum_{k=1,\dotsc,N}\bxi_k\otimes\bzeta_k^\t,\\
\Omega(\bxi,\bxi^*)+\Omega(\bxi,\bxi^*)^\t&=
\sum_{k=1,\dotsc,N}\bxi_k\otimes\bxi_k^\t
\end{aligned}
\end{equation}
 the vectorial fundamental transformation \eqref{vecfun}:
\begin{align*}
\hat\beta_{ij}&=\beta_{ij}-
\langle\bxi^*_j, \Omega(\bxi,\bxi^*)^{-1}\bxi_i\rangle,\\
\hat{\bzeta}_i&=\bzeta_i-\Omega(\bzeta,\bxi^*)\Omega(\bxi,\bxi^*)^{-1}\bxi_i,\\
\hat\bzeta_i^*&=\bzeta_i^*-\bxi_i^*\Omega(\bxi,\bxi^*)^{-1}\Omega(\bxi,\bzeta^*),
\end{align*}
is such that
\[
\hat\bzeta^*_i:=\Big(\frac{\D\hat\bzeta_i}{\D u_i}+
\sum_{\substack{k=1,\dotsc,N\\ k\neq i}}\hat\bzeta_k \hat\beta_{ki}\Big)^{\t}.
\]
\item Moreover, the inversion of \eqref{*l} is
$\bxi_i=\Omega(\bX,\bxi^*)^\t\bX_i$, so that
\[
\Omega(\bX,\bxi^*)=\sum_{k=1}^N\bX_i\otimes\bxi_i^\t.
\]
\end{enumerate}
\end{lem}

The symmetric reduction of the Darboux equations can be combined
with the Lam\'{e} equations to obtain the so called equivalent system
of Darboux-Egoroff equations
\begin{equation}
\label{egor}
\begin{gathered}
\frac{\partial\beta_{ij}}{\D u_k}-\beta_{ik}\beta_{kj}=0,
\;\;
i,j,k=1,\dotsc, N,\;
\text{with $i,j,k$ different},\\
\beta_{ij}-\beta_{ji}=0,\quad i,j=1,\dots,N,\;\; i\neq j,\\
\sum_{k=1}^N\frac{\partial\beta_{ij}}{\D u_k}=0, \quad i,j=1,\dots,N,\;\; i\neq
j.
\end{gathered}
\end{equation}
 This gives rise to the Egoroff metrics; i. e., flat
diagonal metrics \eqref{metric} such that
\eqref{sh} holds.

The reduction of the vectorial fundamental transformation to the
Darboux-Egoroff case can be thought as a superposition of two
reductions, namely the symmetric together with the orthogonal
reduction. Thus, we must request to the transforming data and
potential the constraints for both reductions.  This implies that:
\newtheorem*{RSF}{\textit{\textbf{Vectorial Symmetric Ribaucour Transformation}}}

\begin{RSF}
The vectorial fundamental transformation \eqref{vecfun} preserves
the Darboux-Egoroff equations \eqref{egor}  whenever the
transformation data ($V$, $\bxi_i$, $\bxi_i^*$) satisfy
\begin{gather*}
\bxi_i^*=\bxi_i^\t L=\Big(\frac{\D\bxi_i}{\D u_i}+
\sum_{\substack{k=1,\dotsc,N\\ k\neq i}}\bxi_k \beta_{ki}\Big)^{\t},\\
L^\t\Omega(\bxi,\bxi^*)-\Omega(\bxi,\bxi^*)^\t L=0,\\
\Omega(\bxi,\bxi^*)+\Omega(\bxi,\bxi^*)^\t=
\sum_{k=1,\dotsc,N}\bxi_k\otimes\bxi_k^\t,
\end{gather*}
for some linear operator $L\in\L(V)$.
\end{RSF}

Observe that in the scalar case, $M=1$, the second equation above
is trivial and the third one determines the potential completely.
The associated transformation in this case can be found in
\cite{tsarev}.

\paragraph{Permutability} The vectorial Ribaucour transformation
was shown to have the permutability property in \cite{lm}, moreover
it was done as in our proof of the permutability for the symmetric
case. This implies that the combination of both reductions should
share the permutability character of the symmetric and orthogonal
reduction.

\paragraph{Dressing the Cartesian background}
Now we have:
\[
\bxi_i=\exp(L^\t u_i)\ba_i, \quad \ba_i\in\R^M \text{ constant
vectors,}
\]
and
\[
\Omega(\bxi,\bxi^*)(\bu)=\sum_{i=1,\dotsc, N}\Omega_i(u_i)
\]
 with
\begin{align*}
&
\Omega_i(u_i)=\int_{u_{i,0}}^{u_i}\dif u_i\,\exp(L^\t
u_i)\ba_i\otimes\ba_i^\t\exp(L u_i)L+ \Omega_{i,0} ,\\
&L^\t\Omega_{i,0}-\Omega_{i,0}^\t L=0,\\ &\Omega_{i,0}=
\exp(L^\t
u_{i,0})\ba_i\otimes\ba_i^\t\exp(L u_{i,0})L.
\end{align*}
Thus
\begin{align*}
\Omega(\bX,\bxi^*)(\bu)&=\sum_{i=1,\dotsc, N}
\be_i\otimes\ba_i^\t\exp(L u_i),\\
\Omega(\bxi,H)(\bu)&=\bc+\sum_{i=1,\dotsc, N}
(L^{-1})^\t\exp(L^\t u_i)\ba_i.
\end{align*}
where $\bc\in\R^M$ is a constant vector and we assume that $L$ is
invertible. The  coefficients of the corresponding transformed flat
line element and its flat coordinates are
\begin{align*}
H_i(\bu)^2&=\Big(1-\ba_i^\t\exp(L^\t u_i)L\bigg[\sum_{k=1,\dotsc,
N}\Omega_k(u_k)\bigg]^{-1}\bigg[\bc+
\sum_{l=1,\dotsc, N} (L^{-1})^\t\exp(L^\t u_l)\ba_l\bigg]\Big)^2,\\
 x_i(\bu)&=u_i-
\ba_i^\t\exp(L u_i)\bigg[\sum_{k=1,\dotsc, N}\Omega_k(u_k)\bigg]^{-1}\bigg[\bc+
\sum_{l=1,\dotsc, N} (L^{-1})^\t\exp(L^\t u_l)\ba_l\bigg].
\end{align*}

In the diagonal case, $L=\diag(\ell_1,\dots,\ell_N)$ we find
\begin{align*}
\Omega(\bxi,\bxi^*)&=\big(\Omega_{ij}\big),\quad \Omega_{ij}:=
\frac{\ell_j}{\ell_i+\ell_j}\sum_{k=1}^N\exp((\ell_i+\ell_j)u_k)a_{k,i}a_{k,j},\\
H_i(\bu)^2&=\bigg(1-\dfrac{1}{|\Omega(\bu)|}\sum_{k,l=1,\dots,N}a_{i,k}\exp(\ell_k
u_i)\ell_k\operatorname{cofac}(\Omega(\bu))_{kl}\Big(c_l+
\dfrac{1}{\ell_l}\sum_{j=1,\dotsc, N} \exp(\ell_lu_j)a_{j,l}\Big)\bigg)^2,\\
 x_i(\bu)&=u_i-\dfrac{1}{|\Omega(\bu)|}\sum_{k,l=1,\dots,N}a_{i,k}\exp(\ell_k
u_i)\operatorname{cofac}(\Omega(\bu))_{kl}\Big(c_l+
\dfrac{1}{\ell_l}\sum_{j=1,\dotsc, N}
\exp(\ell_lu_j)a_{j,l}\Big),\\
\beta_{ij}(\bu)&=-\dfrac{1}{|\Omega(\bu)|}\sum_{k,l=1,\dots,N}
\ell_k a_{j,k}\operatorname{cofac}(\Omega(\bu))_{kl}a_{i,l}\exp(\ell_ku_j+\ell_lu_i)
\end{align*}

Where $\ba_l^\t:=(a_{l,1},\dots,\ba_{l,N})$, and we are using the
cofactor matrix $\operatorname{cofac}(A)$; i. e.
$A^{-1}=|A|^{-1}\operatorname{cofac}A$. These type of solutions are
the extension to multidimensions of the bright multi-soliton
solutions of the atractive Nonlinear Schr\"{o}dinger equation, which
describes the propagation of optical pulses in nonlinear fibres.

\end{document}